\documentclass[12pt]{article}

\usepackage{graphicx}
\usepackage{dcolumn}
\usepackage{bm}
\usepackage{soul}
\usepackage{ulem}
\usepackage{color}
\usepackage[T1]{fontenc}
\usepackage[super]{nth}
\usepackage{amsmath}
\usepackage[center]{subfigure}
\usepackage[utf8]{inputenc}
\usepackage{array}
\usepackage{tabu}
\usepackage[super]{nth}
\usepackage{tikz}
\usepackage{standalone}
\usepackage{amssymb}
\usepackage{cite}
\usepackage{float}

\def\beq{\begin{equation}}
\def\eeq{\end{equation}}
\def\ba{\begin{eqnarray}}
\def\ea{\end{eqnarray}}

\begin{document}

\begin{center}
{\Large{\bf Finite-Range Coulomb Gas Models I: Some Analytical Results}} \\
\ \\
\ \\
by \\
Akhilesh Pandey, Avanish Kumar and Sanjay Puri \\
School of Physical Sciences, Jawaharlal Nehru University, New Delhi -- 110067, India.
\end{center}

\begin{abstract}
Dyson has shown an equivalence between infinite-range Coulomb gas models and classical random matrix ensembles for the study of eigenvalue statistics. In this paper, we introduce finite-range Coulomb gas (FRCG) models as a generalization of the Dyson models with a finite range of eigenvalue interactions. As the range of interaction increases, there is a transition from Poisson statistics to classical random matrix statistics. These models yield new universality classes of random matrix ensembles. They also provide a theoretical framework to study banded random matrices, and dynamical systems whose matrix representation can be written in the form of banded matrices.
\end{abstract}

\newpage

\section{Introduction}
\label{s1}

There has been extensive use of random matrices in many branches of physics as well as in other disciplines. For example, they have found applications in quantum chaotic systems, most significantly complex nuclei, atoms, molecules and mesoscopic systems \cite{cp65,ew67,bff81,bg85,bgs84,mm04,cc95,cb97,gmw98,am00,fh01,br05,hs06}. In recent years, novel applications have emerged in biology \cite{ra06,ass15}, economics \cite{pgr99,lcb99}, and communication engineering \cite{tv04,kp10,cd11}. In these applications, classical random matrix ensembles (uniform circular and Gaussian ensembles) have provided a framework for understanding complex spectra.

Dyson demonstrated that the joint probability distribution (jpd) of eigenvalues of these ensembles are equilibrium states of a Brownian motion model of Coulombic particles interacting via a logarithmic potential \cite{fd62}. The positions of the Brownian particles are identified as the eigenvalues of the random matrix ensemble. In his original papers, Dyson considered (a) Coulombic particles moving on a real line with a harmonic binding (referred to as Gaussian ensembles (GE)), and (b) Coulombic particles moving on the unit circle (referred to as circular ensembles (CE)). In related work, Calogero and Sutherland \cite{vp94} have demonstrated that the corresponding quantum Hamiltonians are exactly solvable.

In an important generalization, Dyson also considered non-harmonic confining potentials on the real line \cite{fd72}. In this paper we will refer to these as \textit{linear ensembles}, viz., ensembles of hermitian matrices. Similar generalizations can be done for circular ensembles, viz., ensembles of unitary matrices. The equilibrium properties of these non-Gaussian ensembles have been studied in detail \cite{gpp03,ppk05}. Similarly non-uniform circular ensembles have also been studied \cite{kp08}. Further, non-equilibrium ensembles have also been studied extensively \cite{kp08,sp97,kp11}.

In the Dyson model and the above extensions, all particles have pairwise Coulombic interactions (i.e., each particle interacts with all other particles). In this paper, we consider the natural generalization to the case where particles have finite-range interactions. We will refer to these as {\it finite-range Coulomb gas} (FRCG) models. Such ensembles have important applications in the study of banded random matrices (BRM). Two specific examples are systems of quantum kicked rotors (QKR) and embedded Gaussian ensembles (EGE). Surprisingly, these generalizations have received very limited attention in the literature. In this (paper I) and its companion paper II, we present detailed analytical and numerical results for FRCG models. A brief account of papers I and II has been published in \cite{pkp17}. Some of the analytical results reported in this paper were obtained earlier by one of the authors \cite{apun} and have been used in \cite{bgs99,jk99,bgs01}.

This paper is organized as follows. In Sec.~\ref{s2}, we review Dyson's Brownian motion models. In Sec.~\ref{s3}, we generalize the Dyson models to obtain FRCG models. In Secs.~\ref{s4} and \ref{s5}, we derive the level density of short-range models for linear and circular ensembles respectively. In Sec.~\ref{s6}, we derive the level density of long-range models for both linear and circular ensembles. In Sec.~\ref{s7}, we show the universality of spectral fluctuations with respect to different binding potentials in the circular and linear cases. In Sec.~\ref{s8}, we give some exact results for spectral properties of short-range models ($d=0, 1$), where $d$ is a parameter quantifying the range of the interaction. In Sec.~\ref{s9}, we present a mean-field approximation for $d>1$. In Sec.~\ref{s10}, we present a detailed study of spectral properties for $d=2$ via an integral-equation approach. In Sec.~\ref{s11}, we describe the corresponding integral equation for $d>2$.  We conclude with a summary and discussion in Sec.~\ref{s12}.

\section{Dyson's Brownian Motion Models}
\label{s2}

In this section we will introduce various models used in this paper. We consider $N$-dimensional random hermitian matrices $A$, which can be symmetric hermitian, complex hermitian, quaternion real hermitian or quaternion real self-dual. These are labeled by the Dyson parameter $\beta=1, 2, 4$ respectively \cite{cp65,mm04}. The matrices are $N$-dimensional in real, complex, and quaternion space. In Secs.~\ref{s2a}, \ref{s2b} and \ref{s2c}, we will review Dyson's formulation of Brownian matrix ensembles.

\subsection{Linear and Circular Ensembles}
\label{s2a}

The joint probability density (jpd) of matrices $A$ in the case of Gaussian random matrix ensembles is given by
\begin{equation} \label{e1}
 P(A) = C\exp(-TrA^2/4v^2),
\end{equation}
where $C$ is the normalization constant (we will generally use $C$, $\tilde{C}$ etc. to denote the normalization constants for various distributions). Each matrix element $A_{jk}^{(\gamma)}$ has $\beta$ distinct `sites', labeled by $\gamma = 0,....,\beta -1$. The matrix elements of $A$ follow independent Gaussian distributions and have mean zero and variance $v^2$ for each of the $\beta$ distinct sites. The ensembles corresponding to Eq.~(\ref{e1}) are Gaussian orthogonal ensemble (GOE), Gaussian unitary ensemble (GUE) and Gaussian symplectic ensemble (GSE) for $\beta = 1,2,4$ respectively. A natural generalization of Eq.~(\ref{e1}) is a non-Gaussian ensemble defined by
\begin{equation} \label{e2}
 P(A) = C\exp(-\beta~TrV(A)),
\end{equation}
where $V(A)$ is a positive-definite function of the matrix $A$.

In a similar fashion, one can define circular ensembles of unitary matrices $U$ (symmetric for $\beta=1$, general for $\beta=2$, and self-dual for $\beta=4$). The jpd of $U$ is
\begin{equation}\label{e3}
P(U)=C\exp(-\beta~Tr V(U)),
\end{equation}
where $V(U)$ is a positive-definite function of $U$ and $U^{\dag}$, e.g., $V(U)= U + U^{\dag}$. The case $V=0$ corresponds to the usual circular ensemble introduced by Dyson \cite{fd62}. We will refer to this as the ``uniform circular ensemble'', as opposed to $V \neq 0$ for the ``non-uniform circular ensemble''. These ensembles are referred to as circular orthogonal ensemble (COE), circular unitary ensemble (CUE), and circular symplectic ensemble (CSE) for $\beta = 1,2,4$, respectively. 

\subsection{Brownian Motion Model for Matrices}
\label{s2b}

Dyson \cite{fd62} introduced a Brownian matrix process $M(\tau)$ (over-damped case, i.e., Smoluchowski process \cite{hr84}), in fictitious time $\tau$, yielding Eqs.~(\ref{e1}) and (\ref{e2}) as equilibrium densities. For linear ensembles, $M(\tau)$ represents a Hamiltonian operator. Remarkably this leads to a Brownian process for the eigenvalues of $\lbrace A \rbrace \equiv \lbrace M(\infty) \rbrace$, which interact via a Coulomb gas (logarithmic) potential. In a similar fashion, a Brownian model can be written for the circular case, which yields the equilibrium density in Eq.~(\ref{e3}) \cite{fd62}.

We review Dyson's formulation, starting with the Langevin equation for the matrix variable $M(\tau)$ in the linear ensembles:
\begin{equation}\label{e4}
\frac{d M(\tau)}{d\tau}= -\beta V'(M(\tau))+\xi (\tau).
\end{equation}
Here, $\xi(\tau)$ is a Gaussian white-noise matrix which is similar in structure to $M$. Note that $M(\tau)$ is a matrix variable executing Brownian motion in the space of matrices $M$. The first two moments of the matrix elements $\xi_{jk}^{(\gamma)}$ are \cite{fd62}
\begin{equation}\label{e5}
\overline{\xi_{jk}^{(\gamma)}(\tau)}=0 ,
\end{equation}
\begin{equation}\label{e6}
\overline{\xi_{ij}^{(\gamma)}(\tau)\xi_{kl}^{(\eta)}(\tau')}= \delta_{\gamma\eta}(\delta_{ik}\delta_{jl} +(2\delta_{\gamma 0}-1)\delta_{il}\delta_{jk})\delta(\tau-\tau').
\end{equation}
Here, the bar denotes the ensemble average.

We discretize the Langevin equation over an infinitesimal time interval $\delta\tau$. The matrix increment, $\delta M\equiv M(\tau + \delta \tau)- M(\tau)$, has a drift term (first term of RHS in Eq.~(\ref{e4})) and a diffusive term (second term of RHS in Eq.~(\ref{e4})). The first term contributes to the average and the second term contributes to the covariance of the matrix elements. The average of  $\delta M$ is given by
\begin{equation}\label{e7}
\overline{ \delta M }=-\beta V'(M)\delta\tau .
\end{equation}
The second moments of elements of $\delta M$ are given by
\begin{equation}\label{e8}
\overline{ \delta M_{ij}^{(\gamma)}\delta M_{kl}^{(\eta)} } =\delta_{\gamma\eta} \left[\delta_{ik}\delta_{jl}+(2\delta_{\gamma 0}-1)\delta_{il}\delta_{jk} \right] \delta\tau.
\end{equation}
In the Gaussian case, $\delta M$ is an infinitesimal GOE, GUE, GSE respectively for $\beta=1,2,4$. The off-diagonal variances are $v^{2} \delta \tau$ in each case. The corresponding Fokker-Planck (FP) equation for the probability distribution of the matrix elements is given in \cite{fd72}. The equilibrium jpd of $M(\infty)$ is the same as in Eq.~(\ref{e2}).

For circular ensembles of matrices $U$, the infinitesimal increment is $\delta U= iS\delta M S^{D}$. Here, $S$ is unitary and $S^{D}$ is the transpose of $S$ for $\beta=1$, hermitian adjoint for $\beta=2$, and quaternion dual for $\beta=4$. 

\subsection{Brownian Motion Formulation for Eigenvalues}
\label{s2c}

We consider the eigenvalues $\lbrace x_{j};j=1,\cdots, N \rbrace $ of matrices $M(\tau)$, which obey the Brownian process on the real line. The increment in eigenvalue $\delta x_{j}= x_{j}(\tau + \delta \tau)-x_{j}(\tau)$ for the matrix increment $\delta M$ can be computed by using second-order perturbation theory. Using Eqs.~(\ref{e7})-(\ref{e8}), the first two moments of $\delta x_{j}$ at fixed time $\tau$, correct upto first-order in $\delta \tau$ are
\begin{equation}\label{e9}
\overline{ \delta x_{j} } = \beta \left[ -V'(x_{j})+\sum_{k\neq j}(x_{k}-x_{j})^{-1} \right]\delta\tau= -\beta \frac{\partial W}{\partial x_{j}}\delta \tau ,
\end{equation}
where
\begin{equation}\label{e10}
W= -\sum_{j<k}\log|x_{k}-x_{j}|+\sum_{j}V(x_{j}),
\end{equation}
and
\begin{equation}\label{e11}
\overline{ \delta x_{j}\delta x_{k} } = 2 \delta_{jk}\delta\tau.
\end{equation}
(See Appendix C in \cite{kp11} for the derivation of moments in Eqs.~(\ref{e9}) and (\ref{e11}).) These moments are also known as ``conditional moments''. 

The Fokker-Planck equation for the jpd $p(x_{1},\cdots, x_{N})$ is
\begin{equation}\label{e12}
\frac{\partial p}{\partial \tau}= \sum_{j=1}^{N} \left[  \frac{\partial^2 p}{\partial x_{j}^{2}}-\beta \frac{\partial}{\partial x_{j}} \left( p \frac{\partial W}{\partial x_{j}}\right)  \right].
\end{equation}
The equilibrium jpd of eigenvalues is 
\begin{align}\label{e13}
\begin{split}
p(x_{1},\cdots,x_{N})={}&C \exp(-\beta W) \\
=&C\prod_{j<k}|x_{j}-x_{k}|^{\beta}\exp\Big(-\beta\sum_{j}V(x_{j})\Big) .
\end{split}
\end{align}
Here, $V(x)$= $x^{2}/4v^{2}$ gives the Gaussian ensemble results.

For circular ensembles, one deals with the eigenangles $\theta_{j}$ instead of real variables $x_{j}$. Using second-order perturbation theory for unitary matrices, Eqs.~(\ref{e9})-(\ref{e11}) are replaced by
\begin{equation}\label{e14}
\overline{ \delta \theta_{j} } = \beta \left[ -V'(\theta_{j})+\frac{1}{2}\sum_{k\neq j}\cot\Big(\frac{\theta_{k}-\theta_{j}}{2}\Big) \right]\delta\tau= -\beta \frac{\partial W}{\partial \theta_{j}}\delta \tau,
\end{equation}
where
\begin{equation}\label{e15}
W(\theta_{j})= -\sum_{j<k}\log \Big\vert\sin\Big(\frac{\theta_{k}-\theta_{j}}{2}\Big)\Big\vert+\sum_{j}V(\theta_{j}),
\end{equation}
and
\begin{equation}\label{e16}
\overline{ \delta \theta_{j}\delta \theta_{k} } = 2 \delta_{jk}\delta\tau.
\end{equation}
In Eqs.~(\ref{e14}) and (\ref{e15}), the potential $V(\theta_{j})=V\left(e^{i\theta_j}\right)$.

The Fokker-Planck equation for the jpd $p(\theta_{1},\cdots,\theta_{N})$ is given by Eq.~(\ref{e12}), with $x_{j}$ replaced by $\theta_{j}$ and $W$ as in Eq.~(\ref{e15}). The equilibrium jpd is 
\begin{align}\label{e17}
\begin{split}
p(\theta_{1},\cdots,\theta_{N})=&C \exp\big(-\beta W\big) \\
=&C\prod_{j<k}|\sin(\theta_{j}-\theta_{k})|^{\beta}\exp\Big(-\beta\sum_{j}V(\theta_{j})\Big) .
\end{split}
\end{align}
Note that the logarithmic potential terms in Eqs.~(\ref{e10}) and (\ref{e15}) correspond to the $2$-dimensional Coulomb potential. This is the reason for the usage of the term ``Coulomb gas''. In the next section, we will generalize the above formulation for the FRCG.

\section{{ Finite-Range Coulomb Gas Models}}
\label{s3}

We now introduce FRCG models as a generalization of the above FP equation for the eigenvalue dynamics. In the FP equation (\ref{e12}), we restrict the eigenvalue interaction to the range $d$, i.e., $W$ is given by
\begin{equation}\label{e18}
 W= -{\sum_{j<k}}^{\prime}\log|x_k-x_j| + \sum_{j} V(x_j).
\end{equation}
Here ${\sum}^{\prime}$ denotes the sum over all $|j-k|\leq d$ with $j\neq k$. In Eq.~(\ref{e18}), the logarithmic terms represent the finite-range repulsive two-dimensional Coulomb gas potential, and $V$ is a one-body binding potential. In equilibrium, the jpd of eigenvalues is
\begin{align}\label{e19}
p(x_{1},\cdots,x_{N}) = &C \exp(-\beta W) \nonumber \\ 
=&C{\prod_{j<k}}^{\prime} |x_{j}-x_{k}|^\beta \exp\Big(-\beta\sum_{j}V(x_{j})\Big),
\end{align}
where the prime denotes that the product is restricted to $|j-k|\leq d$. We will refer to these ensembles as \textit{linear ensembles}. Note that we have considered the case of arbitrary potential $V(x)$ in the above discussion. The Gaussian case corresponds to $V(x)= x^{2}/{4v^2}$. 

At this stage, it is appropriate to make some remarks about these FRCG models. \\
1) In the $d=0$ case, the interaction term in Eq.~(\ref{e18}) is absent and the particles $\{x_i\}$ move independently. This corresponds to the Poisson limit. In the $d=N-1$ case, all particles interact with each other. This corresponds to the Wigner-Dyson classical ensembles. Thus, as $d$ changes from 0 to $N-1$, there is a crossover from the Poisson limit to the Wigner-Dyson limit. We would like to understand the nature of this crossover. \\
2) The term ``finite range'' refers to the range in eigenvalue indices, not actual distances. In principle, nearest-neighbor eigenvalues could lie far apart on the real number line. \\
3) It is relevant to ask whether the above crossover is realized in physical systems. In paper II, we will demonstrate that the FRCG models (and their extension to non-integer $d$) provide a framework to understand transitions in QKR and BRM. \\
4) The FRCG models were first proposed by one of the authors \cite{apun}. Their formal properties were studied by Pandey \cite{apun}, Bogomolny et al. \cite{bgs99,bgs01}, and Jain-Khare \cite{jk99}. We will discuss connections to earlier work at appropriate places in papers I and II. The present work constitutes the first detailed application of FRCG models to study transitions in physical systems. \\
5) It is tempting to ask whether the FRCG models can be motivated from a Brownian matrix evolution with banded noise matrices. Then, the $d=N-1$ limit would correspond to the usual Dyson prescription in Secs.~\ref{s2b} and \ref{s2c}. We caution the reader that such a connection is not straightforward. At present, we will treat the FRCG models as generalizations of Dyson's Brownian motion models for eigenvalue spectra. Clearly, an important direction for future work is to identify the matrix ensembles which yield the jpd of eigenvalues in Eq.~(\ref{e19}).

The corresponding banded version of the circular ensemble follows again from Dyson's prescription. The jpd of the resultant equilibrium ensemble is
\begin{equation}\label{e20}
p(\theta_{1}, \cdots,\theta_{N})= C \exp(-\beta W),
\end{equation}
where $\{\theta_j\}$ are eigenangles in ascending order. The term $W$ is given by an appropriate generalization of Eq.~(\ref{e15}):
\begin{equation}\label{e21}
W = - {\sum_{j<k}}^{\prime}\log \Big\vert\sin\Big(\frac{\theta_{k}-\theta_{j}}{2}\Big)\Big\vert + \sum_{j}V(\theta_{j}).
\end{equation}
In this case, $V(\theta)$ is a potential periodic on the unit circle. We can also write $p$ as in Eq.~(\ref{e19}):
\begin{equation}\label{e22}
p(\theta_{1}, \cdots,\theta_{N})= C {\prod_{j<k}}^{\prime} |\sin(\theta_{j}-\theta_{k})|^\beta \exp\Big(-\beta \sum_{j}V(\theta_{j})\Big).
\end{equation}

Note that $V(\theta)=0$ and $d=N-1$ corresponds to the usual Dyson's circular ensemble, which we also refer to as the \textit{uniform circular ensemble}. The introduction of $V$ renders it a \textit{non-uniform circular ensemble} \cite{kp08}. We will concentrate primarily on Gaussian ensembles in the linear case, and uniform ensembles in the circular case. However, we will consider some non-Gaussian and non-uniform circular ensembles also. We will derive analytic results in this paper, and supplement them with Monte Carlo (MC) calculations in paper II. In paper II, we will also discuss applications of FRCG ensembles.

\section{Level Density: Linear Ensembles with $d = O(1)$}
\label{s4}

The level density in the linear case is defined by
\begin{equation}\label{e23}
\rho(x_{1}) = \int_{0}^{\infty} \cdots \int_{0}^{\infty} p(x_{1},\cdots,x_{N})dx_{2} \cdots dx_{N}	.
\end{equation}
The corresponding $p^{\rm th}$ moment is given by
\begin{eqnarray}\label{e24}
M_{p} &=&\int_{0}^{\infty} x^{p}\rho(x)dx \nonumber \\
&= &\int_{0}^{\infty} \cdots \int_{0}^{\infty} x_{j}^{p} p(x_{1},\cdots,x_{N})dx_{1} \cdots dx_{N} \nonumber \\
&=&\frac{1}{N}\sum_{j}\langle x_{j}^{p} \rangle .
\end{eqnarray}
Here, the angular brackets denote
\begin{equation}\label{e25}
\langle F \rangle = \int_{0}^{\infty} \cdots \int_{0}^{\infty} F(x_{1},\cdots,x_{N})p(x_{1},\cdots,x_{N})dx_{1} \cdots dx_{N} ,
\end{equation}
for a function $F(x_{1},\cdots,x_{N})$. After a partial integration, we get
\begin{align}\label{e26}
M_{p}=&~\frac{1}{N}\sum_{j}\bigg\langle \frac{x_{j}^{p+1}}{p+1} e^{-\beta W}\beta \frac{\partial W}{\partial x_{j}} \bigg\rangle \nonumber \\
=&~\frac{\beta}{N(p+1)}\left[-\frac{1}{2}{\sum_{j \ne k}}^{'} \bigg\langle \left(\frac{x_{j}^{p+1} - x_{k}^{p+1}}{x_{j}-x_{k}}\right) \bigg\rangle +\Big \langle \sum_{j} x_{j}^{p+1} V{}^{'}(x_{j})\Big\rangle \right]\nonumber \\
=&~\frac{\beta}{N(p+1)}\left[ -\frac{1}{2}\Big\langle {\sum_{j \ne k}}^{'} \sum_{q=0}^p x_{j}^{q}x_{k}^{p-q} \Big\rangle + \Big\langle   \sum_{j}x_{j}^{p+1} V{}^{'}(x_{j})\Big\rangle\right].
\end{align}
In the second step of the above equation, we have differentiated $W$ and written the double sum in the symmetrized form.

For $d=O(1)$, $x_{j}$ and $x_{k}$ in the double sum of Eq.~(\ref{e26}) can be taken to be equal to each other. For a given $j$, $2d$ values of $k$ contribute to the sum. Thus 
\begin{align}\label{e27}
M_{p} =&~-\beta d M_{p} + \frac{\beta}{N(p+1)} \Big\langle \sum_{j}  x_{j}^{p+1} V{}^{'}(x_{j}) \Big\rangle  \nonumber\\
=&~\frac{\beta}{N(p+1)(\beta d+1)} \Big \langle \sum_{j} x_{j}^{p+1} V{}^{'}(x_{j}) \Big\rangle \nonumber\\
=&~\frac{\beta}{(p+1)(\beta d+1)} \Big \langle x_{j}^{p+1} V{}^{'}(x_{j}) \Big\rangle. 
\end{align}
In the last step of the above equation, $j$ is any of the $N$ indices. Now, using
\begin{equation}\label{e28}
C\int_{0}^{\infty} x^{p} \exp \left(-\frac{\beta V(x)}{\beta d+1}\right)dx= \frac{C \beta}{(p+1)(\beta d+1)}\int_{0}^{\infty} x_{j}^{p+1} V{}^{'}(x_{j})\exp \left(-\frac{\beta V(x)}{\beta d+1}\right),
\end{equation}
we identify the level density as 
\begin{equation}\label{e29}
\rho (x) = C \exp \left(-\frac{\beta V(x)}{\beta d+1}\right).
\end{equation}
Here, $C$ is the normalization constant. Note that, in the Gaussian ensemble, the density has a Gaussian form, but with a larger variance. In paper II, we will verify Eq.~(\ref{e29}) by MC simulations. We will consider the following two potentials: \\
(a) Harmonic potential
\begin{equation}\label{e30}
V(x)= \frac{1}{2}\kappa x^{2}, ~~~~~~~~~\kappa>0.
\end{equation}
(b) Quartic potential
\begin{equation}\label{e31}
V(x)= \kappa\left(\frac{x^{4}}{4}-\alpha\frac{x^2}{2}\right), ~~~~~~\kappa>0.
\end{equation}
In the harmonic and quartic cases, $\kappa$ sets the scale of the $V$-axis. In the quartic potential, $\alpha$ determines whether the potential is single well $(\alpha < 0)$ or double well $(\alpha > 0)$.

\section{Level Density: Circular Ensembles with $d=O(1)$}
\label{s5}

For the circular ensembles, we follow the method described in the previous section. Consider the moment $M_{p}$ of the density $\rho(\theta)$ defined by
\begin{align}\label{e32}
M_{p}=&~\int_{0}^{2\pi} e^{ip\theta}\rho(\theta)d\theta \nonumber \\
=&~ \int_{0}^{2\pi} e^{ip\theta_{j}}p(\theta_{1},\cdots,\theta_{N})d\theta_{1} \cdots d\theta_{N} \nonumber\\
=&~-\frac{C}{N}\sum_{j} \int_{0}^{2\pi} \frac{e^{ip\theta_{j}}}{ip} e^{-\beta W} \left( \frac{\partial{W}}{\partial\theta_{j}}\right)  d\theta_{1} \cdots d\theta_{N} .
\end{align}
Here, a partial integration has been used in the last step. Using $W$ from Eq.~(\ref{e21}), and using angular brackets to denote the ensemble averages as in Eq.~(\ref{e25}), we have
\begin{align}\label{e33}
M_{p}=&~\frac{C}{ipN}\sum_{j} \int_{0}^{2\pi} \frac{e^{ip\theta_{j}}}{ip} e^{-\beta W}\left[ -{\sum_{k}}^{'}\beta \cot\left( \frac{\theta_{j}-\theta_{k}}{2}\right)  + \beta V^{\prime}(\theta)\right]  d\theta_{1} \cdots d\theta_{N} \nonumber \\
=&~\frac{C}{ipN}\left[ -\frac{\beta}{2}~\bigg\langle {\sum_{j \ne k}}^{'}(e^{ip\theta_{j}}-e^{ip\theta_{k}})\cot\left( \frac{\theta_{j}-\theta_{k}}{2}\right)  + \sum_{j}\beta e^{ip\theta_{j}}V^{\prime}(\theta_{j})\bigg\rangle \right] .
\end{align}
In the first step, we have used
\begin{equation}\label{e34}
\frac{d}{d\theta_{j}}\log \left(\sin \bigg|\frac{\theta_{j}-\theta_{k}}{2}\bigg| \right)= \frac{1}{2}\cot \left(\frac{\theta_{j}-\theta_{k}}{2} \right).
\end{equation}
In the second step of Eq.~(\ref{e33}), symmetrization has been done in the double sum. Now, as in the linear case for $d= O(1)$, we take $\theta_{j}$ and $\theta_{k}$ to be equal in the double sum, leading to
\begin{equation}\label{e35}
M_{p}= \frac{\beta}{(p+1)(\beta d +1)}\langle e^{i(p+1)\theta_{j}} V^{\prime}(\theta_{j}) \rangle .
\end{equation}
This yields
\begin{equation}\label{e36}
\rho(\theta)= C \exp \left(-\frac{\beta V(\theta)}{\beta d+1}\right).
\end{equation}

In subsequent discussion, we will consider the following two potentials: \\
(a) Uniform potential: $V(\theta)=0$, \\
(b) Cosine potential: $V(\theta)=\kappa \cos(\theta)$.

\section{Level Density: Linear and Circular Ensembles with $d = O(N)$}
\label{s6}

We first consider linear ensembles in the \textit{long-range case} $(d=O(N))$. The typical pair of $x_{j}$ and $x_{k}$ in the last step of Eq.~(\ref{e26}) can be taken to be independent. Thus, $M_{p}$ can be written as
\begin{align}\label{e37}
M_{p}= \frac{\beta}{N(p+1)}\left[-\frac{1}{2}{\sum_{j \ne k}}^{\prime} \sum_{q=0}^{p}\big\langle x_{j}^{q}\big\rangle \big\langle x_{k}^{p-q}\big \rangle 
+ \Big\langle \sum_{j} x_{j}^{p+1} {V}^{\prime}(x_{j})\Big \rangle \right]. 
\end{align}
The number of distinct terms in ${\sum}^{\prime}$ can be shown to be $N_{d} =Nd - d(d+1)/2$. Thus, for $d= N-1$, we have $N_{d}= N(N-1)/2$ (i.e., all distinct pairs). For $d=(N-1)/2$, we have $N_{d}=(N-1)(3N-1)/8$. We assume that each term in the double sum in $M_{p}$ contributes equally. Thus, since there are $2N_{d}$ pairs of $(j,k)$, we obtain
\begin{align}\label{e38}
M_{p}=&~ \frac{\beta}{p+1}\left[-\frac{2N_{d}}{2N}\sum_{q=0}^{p}\big\langle x_{j}^{q} \big\rangle \big\langle x_{k}^{p-q} \big\rangle + \Big\langle  x_{j}^{p+1} V{}^{'}(x_{j}) \Big\rangle \right] \nonumber  \\
=&~ \frac{\beta}{p+1}\left[-\frac{N\sigma^{2}}{2}\sum_{q=0}^{p}\big\langle x_{j}^{q} \big\rangle \big\langle x_{k}^{p-q} \big\rangle +  \Big\langle x_{j}^{p+1} V{}^{'}(x_{j}) \Big\rangle \right] .
\end{align}
Here,
\begin{equation}\label{e39}
\sigma^{2} = \frac{2N_{d}}{N^{2}} = \gamma(2-\gamma),
\end{equation}
with $d=\gamma N$. In long-range models,
\begin{equation}\label{e40}
V(x)= Nu(x)
\end{equation}
gives a density with finite support (i.e., independent of $N$). Dividing both sides of Eq.~(\ref{e38}) by $N$ and ignoring $O(N^{-1})$ terms in the limit $N \rightarrow \infty$, we get
\begin{equation}\label{e41}
\frac{\sigma^{2}}{2}\sum_{q=0}^{p}M_{q}M_{p-q} = \int_{0}^{\infty} x^{p+1}{u}^{\prime}(x)\rho(x)dx.
\end{equation}

Note that
\begin{eqnarray}\label{e42}
\frac{1}{2}\sum_{q=0}^{p}M_{q}M_{p-q}&=&\frac{1}{2}\int_{0}^{\infty} \int_{0}^{\infty} \frac{x^{p+1}-y^{p+1}}{x-y} \rho(x)\rho(y) dx dy \nonumber \\
&=&\int_{0}^{\infty} \int_{0}^{\infty} x^{p+1} \frac{\rho(x)\rho(y)}{x-y}dx dy.
\end{eqnarray}
Comparing the final steps of Eqs.~(\ref{e41}) and (\ref{e42}), we obtain
\begin{equation}\label{e43}
\sigma^{2}\int_{0}^{\infty} \frac{\rho(y)}{x-y}dy ={u}^{\prime}(x),
\end{equation}
where the principal value of the integral is implied. Integral equations of this type have been considered earlier; see for example \cite{gpp03,gp02}.

As a first example, we consider the above equation for the Gaussian case $u(x)= x^{2}/4$. Then, Eq.~(\ref{e43}) yields the semi-circular density
\begin{equation}\label{e44}
\rho(x)= \frac{\sqrt{4\sigma^{2} - x^{2}}}{2\pi\sigma^{2}}.
\end{equation}
Notice that the semicircle radius $\sigma^{2}=1$ for $d=N-1$, which is the usual Wigner result. For $d \leq N-1$, the semicircle applies once again but with reduced radius. Consider also the Jacobi ensemble, with $V(x)= a\log(1-x)+b\log(1+x)$ where $a$ and $b$ are of order $1$. Since $V$ is independent of $N$, $u(x)=0$ for $|x|< 1$. In this case, the density is independent of $\sigma^{2}$, given by
\begin{equation}\label{e45}
\rho(x)= \frac{1}{\pi\sqrt{1-x^{2}}}.
\end{equation}

For the quartic potential, $u(x)= (\kappa/N) (x^4/4-\alpha x^2/2)$. For $d=N-1$, it is known that the density undergoes a transition from one-band to two-band at the critical point $\alpha_{c}= \sqrt{2N/\kappa}$ \cite{gpp03}. For $d=O(N)$, the pre-factor becomes $\tilde{\kappa}=\kappa/N\sigma^{2}$, and $\alpha_{c}$ changes to $\tilde{\alpha_{c}}= \sigma \alpha_{c}$. The results for the density are
\begin{equation}\label{e46}
\rho(x)= \frac{\tilde{\kappa}}{\pi} \left\{ \frac{1}{3} \left[ \sqrt{\left(\alpha^{2}+\frac{6}{\tilde{\kappa}} \right)}-2\alpha\right] + x^{2} \right\}
\left\{\frac{2}{3}\left[\sqrt{\left(\alpha^{2}+\frac{6}{\tilde{\kappa}} \right)}+\alpha \right] - x^{2}\right\}^{1/2}, \alpha < \tilde{\alpha_{c}},
\end{equation}
and
\begin{equation}\label{e47}
\rho(x)= \frac{\tilde{\kappa}}{\pi}|x| \sqrt{\frac{2}{\tilde{\kappa}}-(x^{2}-\alpha)^{2}}, \quad  \alpha > \tilde{\alpha_{c}}.
\end{equation}
The support of these densities is single-band in Eq.~(\ref{e46}) and two-band in Eq.~(\ref{e47}) respectively.

For $d=O(N)$, with $V(\theta)=N u(\theta)$ in the circular case, we take $\theta_{j}$ and $\theta_{k}$ to be independent as in the linear case, giving thereby
\begin{equation}\label{e48}
\sigma^{2}\int_{0}^{2\pi} \cot \left(\frac{\theta - \phi}{2} \right)\rho(\phi)d\phi = u^{\prime}(\theta).
\end{equation}
Here, again, the principal value of the integral is implied. $\sigma^{2}$ is defined as in Eq.~(\ref{e39}).

When $u(\theta)=0$, the density is uniform, given by
\begin{equation}\label{e49}
\rho(\theta)= \frac{1}{2\pi}.
\end{equation}
Note that this simple result is a consequence of the circular symmetry, and is valid for all $d$ and $N$. This will be useful in computing the fluctuation results for large $d$.

For the non-uniform potential $u(\theta)= \kappa\cos(\theta)$, $\kappa$ changes to $\tilde{\kappa}= \kappa/\sigma^{2}$ for $d=O(N)$. The density is given by \cite{kp08,gw80}
\begin{equation}\label{e50}
\rho(\theta)= \frac{1}{\pi}\left[\kappa(1-\cos\theta) (1-\kappa-\kappa\cos\theta)\right]^{1/2},  \kappa\geq \frac{1}{2},
\end{equation}
and
\begin{equation}\label{e51}
\rho(\theta)= \frac{1}{2\pi} \left[1-2\kappa\cos\theta \right], 0\leq \kappa \leq \frac{1}{2}.
\end{equation}
The density in Eq.~(\ref{e50}) corresponds to the banded case with a peak at $\theta=\pi$. On the other hand, the density in Eq.~(\ref{e51}) describes the non-banded case, valid for the entire range of $\theta$ (again peaked at $\theta=\pi$).

Note that the critical point $\tilde{\alpha_{c}}$ decreases with $d$ in the quartic case, whereas the effective transition parameter $\kappa$ in the circular case increases with $\sigma^{2}$. Thus, in both the cases, the onset of the transition occurs more rapidly with decreasing $d$.

\section{Equivalence of Fluctuations for Linear and Circular Ensembles}
\label{s7}

As we have mentioned earlier, the well-known classical ensembles (Gaussian linear as well as uniform circular) arise for $d=(N-1)$ \cite{mm04}. For these ensembles, it has been proven analytically that the fluctuation properties are identical for each $\beta$ after proper unfolding of the spectra, i.e., there is a distinct universality class for each $\beta$. Further, these properties are also the same for an arbitrary smooth potential $V$ \cite{pkp17,gp02,pg01}.

Therefore, it is natural to investigate the nature of universality for short-range models. In this section, we demonstrate that \\
(a) For $d=O(1)$: Each ($d, \beta$) gives rise to a distinct universality class. \\
(b) For $d=O(N)$: For each $\beta$, the corresponding universality class of classical ensembles applies.

Let us first focus on $d=O(1)$ cases for the circular ensemble. We define
\begin{equation}\label{e52}
s_{j}= \frac{(\theta_{j+1}-\theta_{j})}{D(\theta_{j})},
\end{equation}
where $D(\theta)$ is the average spacing at $\theta $. The $\theta_{j}$'s are defined modulo $2\pi$. Note that the spacing $s_{j}$ has an average equal to $1$ for all $j$. This is commonly referred to as unfolding of the spectra \cite{bff81}. $D(\theta)$ is given by
\begin{equation}\label{e53}
D(\theta) = \frac{1}{N \rho(\theta)},
\end{equation}
where $\rho(\theta)$ is the normalized level density given in Eq.~(\ref{e36}). For example, in the case of zero potential, $D(\theta)= 2\pi/N$. Note that, for large $N$ and $j>k$,
\begin{equation}\label{e54}
\frac{1}{D(\theta)}\Big|2 \sin\frac{(\theta_{j}-\theta_{k})}{2}\Big|= \frac{(\theta_{j}-\theta_{k})}{D(\theta)}=(s_{j-1}+\cdots+ s_{k}).
\end{equation}
Thus, using Eq.~(\ref{e22}), we obtain the jpd of the nearest neighbor spacings $s_{j}$ as
\begin{equation}\label{e55}
P_{d}(s_{1},\cdots,s_{N})=C_{d}\delta\left(\sum_{i=1}^{N} s_i-N\right) 
\prod_{j=1}^{N}[ s_{j}.(s_{j}+s_{j+1})\cdots (s_{j}+\cdots + s_{j+d-1})]^{\beta},
\end{equation}
valid for $N\gg d \geq 1$. This can also be written as
\begin{align}\label{e56}
P_{d}&=C_{0}\delta\bigg(\sum_{i=1}^{N}s_{i}-N\bigg),&d=0,				\nonumber \\
&=C_{d}\delta\bigg(\sum_{i=1}^{N}s_{i}-N \bigg) \prod_{j=1}^{N}\prod_{k=0}^{d-1}(s_{j}+\cdots+s_{j+k})^{\beta},&d\geq1.
\end{align}
The $\delta$-function term in Eqs.~(\ref{e55}), (\ref{e56}) arises because the eigenvalues lie on the unit circle. Note that, in deriving Eqs.~(\ref{e55}) and (\ref{e56}), we have used Eq.~(\ref{e36}) to obtain
\begin{equation}\label{e58}
\bigg[ \prod_{j}\rho(\theta_{j})\bigg]  ^{\beta d +1}= e^{-\beta \sum_{j} V\left( \theta_{j}\right)}.
\end{equation}
Thus, the entire $V$-dependence is eliminated in the $\lbrace \theta_{j} \rbrace$  $\rightarrow$ $\lbrace s_{j}\rbrace$ transformation, including the contribution from the Jacobian of the transformation. Therefore, the fluctuation properties are independent of $V(\theta)$ for $d=O(1)$ in the circular ensemble.

In a similar way, for the linear ensembles we use Eqs.~(\ref{e52}) and (\ref{e53}) with $\theta$ replaced by $x$. In this case, the density $\rho(x)$ is given by Eq.~(\ref{e29}). Now, using Eq.~(\ref{e19}), we obtain Eqs.~(\ref{e55}) and (\ref{e56}) without the $\delta$-function term. For large $N$, the mean of  $\sum s_{j}$ is $N$ with relative fluctuations of $O(N^{-1/2})$. Therefore, the $\delta$-function term can be reinstated. This proves the equivalence of circular and linear ensembles for the fluctuation properties for large $N$ with $d=O(1)$. This result is independent of the potential $V$. 

Let us next turn to the case of $d=O(N)$. Eqs.~(\ref{e55}) and (\ref{e56}) are valid for large $N$. To take the $N\rightarrow \infty$ limit, we need to consider the jpd of $n$ consecutive spacings where $n \ll N$. Thus,
\begin{equation}\label{e59}
P^{(n)}_{d}(s_{1},\cdot\cdot, s_{n}) \equiv \lim_{N\to\infty}  \int_{0}^{\infty}\cdot\cdot\int_{0}^{\infty}P_{d}(s_{1},\cdot\cdot,s_{N})ds_{n+1}\cdot\cdot ds_{N} 
\end{equation}
is the jpd for each $\beta$ and $d$, and independent of $V$. We will deal with explicit forms of $P^{(n)}_{d}$ in sections \ref{s8}, \ref{s9} and \ref{s10}. 

The potential $V$ is important in determining the density; see Eqs.~(\ref{e29}) and (\ref{e36}). On the other hand, the fluctuations on a scale $n \ll N$ are governed by the logarithmic potential. When $d=O(N)$ it is natural to expect that, after proper unfolding, the fluctuations on the scale of order $n$ will be independent of $d$ and therefore universal for each $\beta$.

The above discussion demonstrates the equivalence of linear and circular ensembles. Therefore, in our subsequent discussion, we will focus on the case of circular ensembles with potential $V(\theta)=0$.

We now introduce the statistical measures used to characterize eigenvalue fluctuations. These are as follows \cite{bff81,mm04}. \\
\\
1) $(n-1)^{\rm th}$ nearest neighbor spacing distribution $(n=1,2,3,\cdots)$
\begin{align}\label{e60}
p_{n-1}(s)=\int_{0}^{\infty}\cdots\int_{0}^{\infty} \delta \left(s-\sum_{j=1}^{n}s_{j}\right) \times \nonumber\\
P_{d}^{(n)}(s_{1},\cdots s_{n})ds_{1}\cdots ds_{n}.
\end{align}
A related quantity is
\begin{equation}\label{e61}
Q_{n}(s_{1},s_{n})=\int_{0}^{\infty}\cdots\int_{0}^{\infty}ds_{2}\cdots ds_{n-1}P_{d}^{(n)}(s_{1},\cdots s_{n}),
\end{equation}
which is the jpd of two separated spacings $s_{1}$ and $s_{n}$. \\
\\
2) Spacing variance
\begin{equation}\label{e62}
\sigma^{2}(n-1)= \int_{0}^{\infty} s^{2}p_{n-1}(s)ds - n^{2}.
\end{equation}
3) Two level correlation function $R_{2}$ and cluster function $Y_{2}$.
\begin{equation}\label{e63}
R_{2}(s)= 1- Y_{2}(s)= \sum_{n=1}^{\infty}p_{n-1}(s).
\end{equation}
4) Number variance
\begin{equation}\label{e64}
\Sigma{}^{2}(r)= r-2\int_{0}^{\infty}(r-s)Y_{2}(s)ds.
\end{equation}

Note that the results for $d=(N-1)$ with $N\rightarrow \infty$ correspond to the standard results given by Dyson, Mehta and others \cite{cp65,mm04}. In Wigner-Dyson statistics, all the results for eigenvalue and eigenvector fluctuations correspond to the $N\rightarrow \infty$ limit. In paper II, we will demonstrate from MC calculations that these results are applicable for $d=O(N)$. 
Here, we briefly review the main analytical results for fluctuation properties. These will be used for comparison with (a) results from numerical integration presented in this paper, and (b) the MC results presented in paper II.

The exact spacing distributions have complicated analytical forms deriving from integral equations. However, $p_0(s)$ is well-approximated by the following results \cite{mm04}:
\begin{align}
p_{0}(s)&=\frac{\pi}{2}s\exp \left(-\frac{\pi}{4}s^{2}\right), \quad \beta=1,  	\label{e65} \\
p_{0}(s)&=\frac{32}{\pi^{2}}s^{2}\exp \left(-\frac{4}{\pi}s^{2} \right), \quad \beta=2,	 \label{e66} \\
p_{0}(s)&=\frac{2^{18}}{3^{6}\pi^{3}}s^{4}\exp \left(-\frac{64}{9\pi}s^{2}\right), \quad \beta=4. \label{e67}
\end{align}

The number variance for $r \gtrsim 1$ is given by
\begin{equation}\label{e68}
\Sigma{}^{2}(r)= \frac{2}{\beta\pi^{2}}\ln r + c_{\beta},
\end{equation}
where 
\begin{align}
c_{1}&=\frac{2}{\pi^{2}} \left[\ln(2\pi)+\gamma +1 -\frac{\pi^{2}}{8} \right],  \label{e69} \\
c_{2}&=\frac{1}{\pi^{2}} \left[\ln(2\pi)+\gamma +1 \right],					 \label{e70}  \\
c_{4}&=\frac{1}{2\pi^{2}} \left[\ln(4\pi)+\gamma +1 + \frac{\pi^{2}}{8} \right]. \label{e71}
\end{align}
Here, $\gamma = 0.5772156$ is the Euler constant. The spacing variance (for $\beta=1,2,4$) up to a good approximation is given by \cite{fmp78,bff81}
\begin{equation}\label{e72}
\Sigma{}^{2}(k+1)=\sigma^{2}(k)+\frac{1}{6},
\end{equation}
which is exact for $k\rightarrow \infty$.

The two-level cluster functions are known exactly \cite{mm04}:
\begin{align}
Y_{2}(r)&=\left(s(r)\right)^{2}+\left(\int_{r}^{\infty}s(t)dt\right)\left(\frac{d}{dr}s(r)\right), \quad \beta=1, \label{e73}\\
Y_{2}(r)&= \left(s(r)\right)^{2} \label{e74}, \quad \beta=2,	\\
Y_{2}(r)&=\left(s(2r)\right)^{2}+\left(\int_{0}^{r}s(2t)dt\right)\left(\frac{d}{dr}s(2r)\right), \quad \beta=4 . \label{e75}
\end{align}
Here,
\begin{equation}\label{e76}
s(x)= \frac{\sin(\pi x)}{\pi x}.
\end{equation}

We will see in paper II that the Dyson-Mehta results already arise for FRCG models with large values of $d$, even for $d\ll N$.

\section{Fluctuation Properties for $d = 0,1$}
\label{s8}

We start with the simplest case, viz., $d=0$. From Eq.~(\ref{e59})
\begin{align}\label{e77}
&P_{0}^{(n)}(s_{1},\cdots,s_{n})\nonumber \\
&= \lim_{N\to\infty} C_{0} \int_{0}^{\infty} \delta \left(\sum_{j=1}^{N}s_{j}-N \right) ds_{n+1}\cdots ds_{N} \nonumber \\
&=\lim_{N\to\infty} \tilde{C}_{0} \left(N-\sum_{j=1}^{n}s_{j}\right)^{N-n-1} \nonumber \\
&=	\prod_{j=1}^{n} e^{-s_{j}}.
\end{align}
In the first step of Eq.~(\ref{e77}), the integration is facilitated by the change of variables 
\begin{equation}
x_{j}= \frac{s_j}{(N-\sum_{i=1}^{n}s_{i})}, \quad j=n+1,\cdots N,
\end{equation}
and using $\delta (ax)=|a|^{-1}\delta(x)$. In the last step, the normalization constant is unity. This is the standard Poisson result as there is no interaction between the ``particles''. The spacings $s_{j}$ are independent and have exponential distributions. Note that the results in this case are independent of $\beta$.

In this case, the fluctuation measures given in Eqs.~(\ref{e60})-(\ref{e64}) yield the following well-known results:
\begin{align} \label{e78}
p_{n-1}(s)&=\frac{s^{(n-1)}}{(n-1)!}e^{-s}, & \sigma^{2}(n-1)&=n, 
\end{align}
and
\begin{align}\label{e79}
R_{2}(s)&=1,& Y_{2}(s)&=0, & \Sigma{}^{2}(r)&=r.			
\end{align}

Next, we consider the $d=1$ case. Now, each particle (i.e., eigenvalue) interacts with its nearest neighbor. From Eq.~(\ref{e56}) it is clear that there will be an extra factor $s_{j}^{\beta}$ in the jpd $P_{1}(s_{1}\cdots s_{N})$. Again, from Eqs.~(\ref{e55}) and (\ref{e59}), we find
\begin{align}\label{e80}
P_{1}^{(n)}(s_{1},\cdots s_{n})=&~\lim_{N\to\infty} C_{1} \int_{0}^{\infty} \prod_{j=1}^{N}s_{j}^{\beta}\delta \left(\sum_{i=1}^{N} s_i-N \right) ds_{n+1}\cdots ds_{N}\nonumber \\
=&~\lim_{N\to\infty}\tilde{C_{1}} \Big(N-\sum_{i=1}^{n}s_{i}\Big)^{(N-n)(\beta+1)}\prod_{j=1}^{n}s_{j}^{\beta}\nonumber\\
=&~\prod_{j=1}^{n} \frac{(\beta+1)^{\beta+1}}{\beta !}s_{j}^{\beta}e^{-(\beta+1)s_{j}}.
\end{align}

In this case, the spacings are again independent, and
\begin{equation}\label{e81}
p_{0}(s)= \frac{(\beta+1)^{(\beta+1)}}{\beta!}s^{\beta} e^{-(\beta+1)s}.
\end{equation}
Further, the  $(n-1)^{\rm th}$  spacing distribution is given by
\begin{equation}\label{e82}
p_{n-1}(s)=\frac{(\beta+1)^{(\beta+1)n}}{((\beta+1)n-1)!} s^{(\beta+1)n-1}e^{-(\beta+1)s}.
\end{equation}
The proof of this also follows directly from Eq.~(\ref{e60}) by change of variables $s_{j}=s x_{j}$. The spacing variance is obtained as
\begin{equation}\label{e83}
\sigma^{2}(n-1)= \frac{n}{\beta+1}.
\end{equation}

To obtain the number variance $\Sigma^2$, we first consider the Laplace transform of the two-level cluster function. Using Eqs.~(\ref{e63}) and (\ref{e82}), and introducing $\alpha$ as the Laplace variable, we find
\begin{align}\label{e84}
\frac{1}{\alpha} - \int_{0}^{\infty} e^{-\alpha s}Y_{2}(s)ds=&\left(\frac{\beta+1}{\beta+1+\alpha}\right)^{\beta+1}\frac{1}{1-\left(\frac{\beta+1}{\beta+1+\alpha}\right)^{(\beta+1)}} \nonumber \\
=&\frac{1}{\alpha} - \frac{\beta}{2(\beta+1)}+\frac{\beta(\beta+2)}{12(\beta+1)^{2}}\alpha + O(\alpha^{2}).
\end{align}
Comparing the constant term, and the $\alpha$-order term, we have
\begin{equation}\label{e85}
\int_{0}^{\infty}Y_{2}(s)ds = \frac{\beta}{2(\beta+1)},
\end{equation}
\begin{equation}\label{e86}
 \int_{0}^{\infty}s Y_{2}(s)ds = \frac{\beta(\beta+2)}{12(\beta+1)^{2}}.
\end{equation}
Since the two $Y_{2}$ integrals in Eq.~(\ref{e64}) converge rapidly, we can use Eqs.~(\ref{e85}) and (\ref{e86}) in Eq.~(\ref{e64}) to get
\begin{equation}\label{e87}
\Sigma{}^{2}(r)=\frac{r}{\beta+1} + \frac{\beta(\beta+2)}{6(\beta+1)^{2}}.
\end{equation}

The rapid convergence of integrals comes from the fact that the inverse Laplace transform gives exponential terms in $Y_{2}$. For example, for $\beta=1$ we find
\begin{equation}\label{e88}
Y_{2}(s)= e^{-4s}.
\end{equation}
For other $\beta$-values, $Y_{2}(s)$ contains several exponential terms and is given in Eq.~(\ref{e95}) for $d=1, \beta=2$.

\section{Mean-Field Approximation for $d>1$}
\label{s9}

In this section, we present an approximation which reduces the arbitrary $d$ case to an effective $d=1$ case. This reduction is analogous to the mean-field (MF) approximation in statistical mechanics. The MF approximation will give good estimates for the above fluctuation measures. In Eq.~(\ref{e55}), we set $s_{j+1},s_{j+2},\cdots, s_{j+d-1}\simeq s_{j}$ in each of the factors under the product sign, neglecting fluctuations in neighboring spacings. This yields
\begin{equation}\label{e89}
P_{d}(s_{1},\cdots,s_{N})\simeq \bar{C}_{d}\delta\bigg(\sum_{i=1}^{N}s_{i}-N \bigg) \prod_{j=1}^{N}s_{j}^{\beta d}.
\end{equation}
Then, integrating over variables $s_{n+1},\cdots,s_{N}$ as in the $d=0,1$ cases, we obtain
\begin{align}\label{e90}
P_{d}^{(n)}\simeq \lim_{N\to\infty}\tilde{C_{d}} \Big(N-\sum_{i=1}^{n}s_{i}\Big)^{(N-n)\xi}\prod_{j=1}^{n}s_{j}^{\beta d}\nonumber \\
= \prod_{j=1}^{n} \frac{\xi^{\xi}}{\Gamma(\xi)}s_{j}^{(\xi-1)}e^{-\xi s_{j}}, ~~ \xi= (\beta d+1),
\end{align}
i.e., the spacings are independent for large $N$. This approximation is valid for $n  \gtrsim d$. Note that Eq.~(\ref{e90}) is exact for $d=0,1$.

Substituting Eq.~(\ref{e90}) in Eq.~(\ref{e60}), we obtain the spacing distribution for $n \geq 1$:
\begin{equation}\label{e91}
p_{n-1}(s)= \frac{\xi^{n\xi}}{\Gamma(n\xi)} s^{n\xi -1}e^{-\xi s}.
\end{equation}
The spacing variance is given by
\begin{equation}\label{e92}
\sigma^{2}(n-1)=\frac{n}{\beta d +1} + \gamma(\beta,d).
\end{equation}
The leading term in Eq.~(\ref{e92}) (i.e., the linear term in $n$) is determined by the MF result in Eq.~(\ref{e91}). The constant term $\gamma(\beta,d)$ does not arise in the MF approximation, and its introduction is motivated by our exact $d=2$ result below. For $d>2$, we estimate this constant from our MC calculations as discussed in paper II.

To calculate the number variance, we use Eqs.~(\ref{e90}) and (\ref{e91}) to obtain the Laplace transform of the two-level correlation function $R_{2}(s)$ in Eq.~(\ref{e63}). We find for $\alpha\geq 0$,
\begin{align}\label{e93}
\frac{1}{\alpha}-\int_{0}^{\infty}e^{-\alpha s}Y_{2}(s)ds =&~\sum_{n=1}^{\infty}\frac{1}{\left[ (1+\frac{\alpha}{\xi})^{\xi}\right]^{n}} \nonumber \\
=&~\frac{{\xi}^{\xi}}{\left[ ({\xi+\alpha})^{\xi}-{\xi}^{\xi}\right]} .
\end{align}
Here, the final expression results from the sum of an infinite geometric series. The cluster function $Y_{2}$ falls off exponentially for all $d=O(1)$. For example, for $d=1, \beta=2$ (i.e., $\xi=3$), we have from Eq.~(\ref{e93}) after taking the inverse Laplace transform:
\begin{equation}\label{e94}
\mathcal{L}^{-1}\left\lbrace \int_{0}^{\infty}e^{-\alpha s}Y_{2}(s)ds\right\rbrace =\mathcal{L}^{-1} \left\lbrace  \frac{1}{\alpha}-\frac{27}{(3+\alpha)^{3}-27}\right\rbrace .
\end{equation}
This gives 
\begin{equation}\label{e95}
Y_{2}(s)= 2 e^{-9s/2} \sin\left( \frac{3}{2}\sqrt{3}s + \frac{\pi}{6}\right).
\end{equation}
Using Eq.~(\ref{e93}), we can evaluate the number variance $\Sigma^{2}(r)$ in Eq.~(\ref{e64}) as
\begin{equation} \label{e96}
\Sigma^{2}(r)= \sigma^{2}(r-1) + \frac{(\xi^2 -1)}{6 \xi^{2}} .
\end{equation}

For $d>2$, one can improve the MF approximation by considering a reduction to an effective $d=2$ case. This process gives an estimate for $\gamma(\beta, d)$ in Eq.~(\ref{e92}), but the corresponding calculation requires solutions of the integral equation discussed below for $d=2$.

\section{Fluctuation Properties for $d=2$}
\label{s10}

Let us now consider FRCG ensembles with next-nearest-neighbor interactions, i.e., $d=2$. This case will involve significant complications as compared to the $d=1$ case. This is due to the correlations introduced between the consecutive level spacings. From Eq.~(\ref{e55}), the jpd of the nearest-neighbor spacings for $d=2$ is
\begin{equation}\label{e97}
P_{2}(s_{1},\cdots s_{N})= C_{2}\delta \left(\sum_{i=1}^{N} s_i-N \right)\prod_{j=1}^{N}s_{j}^{\beta}(s_{j}+s_{j+1})^{\beta}.
\end{equation}
The corresponding jpd of $n \geq 2$ consecutive spacings is
\begin{align}\label{e98}
P_{2}^{(n)}(s_{1},\cdots s_{n})=\lim_{N\to\infty}C_{2}\int_{0}^{\infty} \delta \left(\sum_{i=1}^{N} s_i-N \right) \prod_{j=1}^{N}s_{j}^{\beta}(s_{j}+s_{j+1})^{\beta}ds_{n+1}\cdots ds_{N}, 
\end{align}
or
\begin{align}\label{e99}
P_{2}^{(n)}(s_{1},\cdots s_{n})=\tilde{C}_{2} \exp\left[-(2\beta+1)\sum_{i=1}^{n}s_{i}\right]
\prod_{j=1}^{n}s_{j}^{\beta}\prod_{k=1}^{n-1}(s_{k}+s_{k+1})^{\beta}G(s_{1},s_{n}).
\end{align}
Here, the exponential factor arises from the $\delta$-function. The factor $(2\beta+1)$ becomes $(\beta d+1)$ for general $d$, as there are $d$ multiplying factors of $s$-terms. The term $G(s_{1},s_{n})$ arises because there are two extra factors in Eq.~(\ref{e97}) which depend on $s_{1}$ and $s_{n}$, viz., $(s_{1}+s_{N})^{\beta}$ and $(s_{n}+s_{n+1})^{\beta}$. 

As $s_{1}$ and $s_{n}$ are well-separated by the integration variables $(s_{n+1},\cdots,s_{N})$, and the interaction is short-ranged, we assume that $G(s_{1},s_{n})$ is factorisable:
\begin{equation}\label{e100}
G(s_{1},s_{n}) \simeq F(s_{1})F(s_{n}).
\end{equation}
From Eq.~(\ref{e99}), a further integration on $s_{n}$ yields $P_{2}^{(n-1)}(s_{1},\cdots,s_{n-1})$. In this process, the function $F(s)$ does not change, but the constant changes. This implies that $F(s)$ is an eigenfunction of the integral equation:
\begin{equation}\label{e101}
\int_{0}^{\infty}e^{-(2\beta+1)s}s^{\beta}(t+s)^{\beta}f_{\mu}(s)ds=\lambda_{\mu}f_{\mu}(t).
\end{equation}
Here, $\lambda_{\mu}$ and $f_{\mu}$ are eigenvalues and eigenfunctions of this equation. The kernel $(t+s)^{\beta}$ is a polynomial of order $\beta$ in $t$, implying that there are $(\beta+1)$ eigenvalues. We order them as $\lambda_{0}<\lambda_{1}\cdots<\lambda_{\beta}$. The eigenfunction $f_{\mu}$ is a polynomial of order $\beta$, and this enables us to solve Eq.~(\ref{e101}) numerically by matrix diagonalization. The eigenfunction $f_{\beta}$ corresponding to the largest eigenvalue will have only positive coefficients. Thus, for $f_{\beta}$ normalized as in Eq.~(\ref{e103}) below, we obtain
\begin{align}\label{e102}
P_{2}^{(n)}(s_{1},\cdots s_{n})= \frac{1}{\lambda_{\beta}^{n-1}}\exp\left[-(2\beta+1)\sum_{j=1}^{n}s_{j}\right] \times \nonumber \\
\prod_{j=1}^{n}s_{j}^{\beta}\prod_{k=1}^{n-1}(s_{k}+s_{k+1})^{\beta}f_{\beta}(s_{1})f_{\beta}(s_{n}),
\end{align}
valid for $n=2,3,\cdots$.

It is useful to describe some properties of the integral equation (\ref{e101}). The kernel $(t+s)^{\beta}$ is symmetric in $t$ and $s$, so the integral equation is hermitian. Therefore, the eigenvalues $\lambda_{\mu}$ are real and the eigenfunctions $f_{\mu}$ are orthogonal. The eigenfunctions are appropriately normalized, giving the orthonormality relation:
\begin{equation}\label{e103}
\int_{0}^{\infty}e^{-(2\beta+1)s} s^{\beta}f_{\mu}(s)f_{\nu}(s)ds= \delta_{\mu\nu}.
\end{equation}
The properties of hermitian operators can be exploited to obtain sum rules involving the eigenvalues and eigenfunctions. The kernel $I(s,t)$ of the identity operator can be expressed as 
\begin{equation}\label{e104}
I(s,t)= \sum_{\mu =0}^{\beta}f_{\mu}(s)f_{\mu}(t).
\end{equation}
The \textit{spectral decomposition} property can be written as
\begin{equation}\label{e105}
(t+s)^{\beta}= \sum_{\mu =0}^{\beta}\lambda_{\mu}f_{\mu}(s)f_{\mu}(t).
\end{equation}
This property can be generalized, using Eqs.~(\ref{e101}) and (\ref{e104}), as 
\begin{align}\label{e106}
&\int_{0}^{\infty}\cdots\int_{0}^{\infty} ds_{2}\cdots ds_{n} \exp\left[-(2\beta+1)\sum_{j=1}^{n}s_{j}\right]\times \nonumber\\
&\prod_{k=1}^{n}s_{k}^{\beta}(s_{1}+s_{2})^{\beta}
(s_{2}+s_{3})^{\beta}\cdots(s_{n}+s_{n+1})^{\beta}\nonumber\\
&=\sum_{\mu=0}^{\beta}\lambda_{\mu}^{n}f_{\mu}(s_{1})f_{\mu}(s_{n+1}).
\end{align} 
Notice that, for $n=1$ in Eq.~(\ref{e106}), we recover Eq.~(\ref{e105}). Finally, we have the trace relation
\begin{align}\label{e107}
&\int_{0}^{\infty}\cdots\int_{0}^{\infty} ds_{1}\cdots ds_{n}\exp\left[-(2\beta+1)\sum_{j=1}^{n}s_{j}\right]\times \nonumber\\
&\prod_{k=1}^{n}s_{k}^{\beta}(s_{1}+s_{2})^{\beta}
(s_{2}+s_{3})^{\beta}\cdots(s_{n}+s_{1})^{\beta}\nonumber\\
&=\sum_{\mu=0}^{\beta}\lambda_{\mu}^{n}.
\end{align}

In Eq.~(\ref{e102}), we integrate over all variables except $s_{1}$ and $s_{n}$ to obtain
\begin{align}\label{e108}
Q_{n}(s_{1}, s_{n})= e^{-(2\beta+1)(s_{1}+s_{n})}s_{1}^{\beta}s_{n}^{\beta}f_{\beta}(s_{1})f_{\beta}(s_{n})
\sum_{\mu=0}^{\beta}\left( \frac{\lambda_{\mu}}{\lambda_{\beta}}\right) ^{n-1}f_{\mu}(s_{1})f_{\mu}(s_{n}).
\end{align}
For $n=1$, this gives the nearest-neighbor spacing distribution
\begin{equation}\label{e109}
p_{0}(s)= Q_{1}(s_{1})= e^{-(2\beta+1)s}s^{\beta}\left[f_{\beta}(s)\right]^{2}.
\end{equation}
We point out that the distributions in Eqs.~(\ref{e108}) and (\ref{e109}) are normalized to unity. Further, because of unfolding, the average spacings are all unity. 

\begin{figure}[H]
\centering
\includegraphics[width=0.6\textwidth]{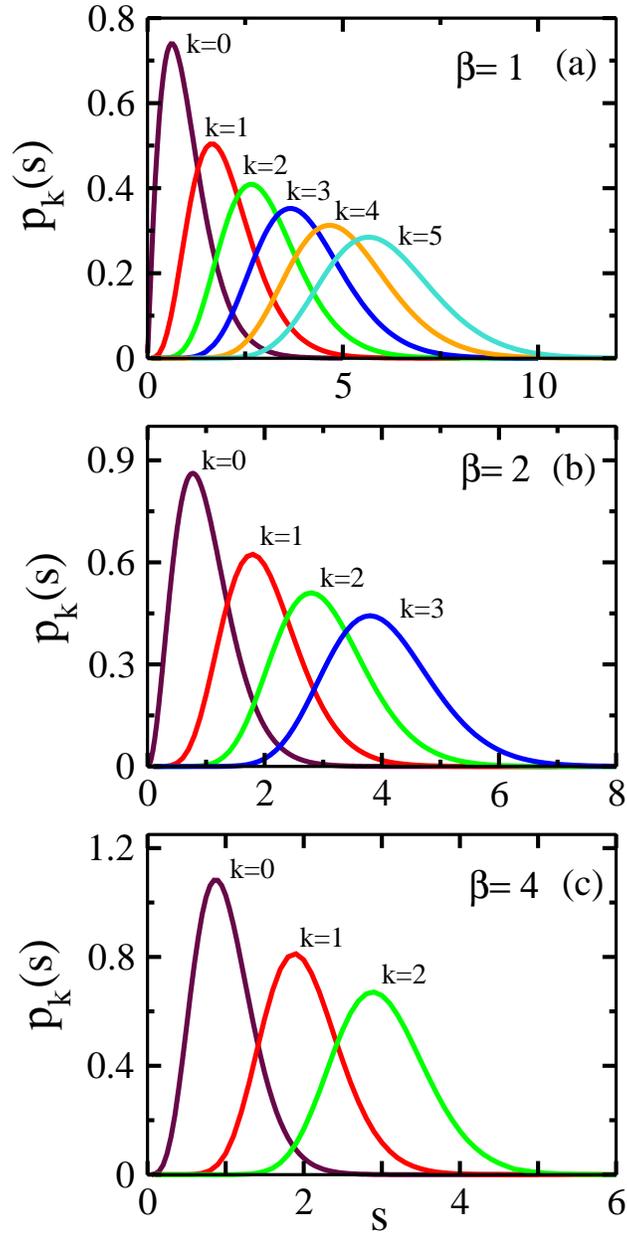}
\caption{Spacing distributions, $p_k(s)$ vs. $s$, for $d=2$. Here, $k$ is the order of the distribution, with $k=0$ corresponding to the nearest neighbor. The different frames correspond to (a) $\beta=1$, (b) $\beta=2$, and (c) $\beta=4$. In each case, we show $p_k(s)$ up to $k=k_m(\beta)$. For $k \geq k_m$, the exact result is numerically indistinguishable from the MF result on this scale.}\label{F1}
\end{figure}

For higher-order spacing distributions $(p_{n-1}$ or $p_{k}$ for $k=n-1, n \geq 1)$, we substitute Eq.~(\ref{e102}) in Eq.~(\ref{e60}) to obtain
\begin{align}\label{e110}
p_{n-1}(s)=p_{k}(s)=\frac{1}{\lambda_{\beta}^{n-1}}\int_{0}^{\infty}\cdots\int_{0}^{\infty} \delta \left(s-\sum_{i=1}^{n}s_{i}\right)\exp\left(-(2\beta+1)\sum_{i=1}^{n}s_{i}\right)  \times \nonumber\\
\prod_{i=1}^{n}s_{i}^{\beta}\prod_{j=1}^{n-1}(s_{j}+s_{j+1})^{\beta}f_{\beta}(s_{1})f_{\beta}(s_{n})ds_{1}\cdots ds_{n}.
\end{align}
Thus, for example,
\begin{equation}\label{e111}
p_{1}(s)=\frac{1}{\lambda_{\beta}}e^{-(2\beta+1)s}\int_{0}^{\infty}s_{1}^{\beta}(s-s_{1})^{\beta}s^{\beta}f_{\beta}(s_{1})f_{\beta}(s-s_{1})ds_{1}.
\end{equation}
We have performed a numerical integration of Eq.~(\ref{e110}), and plot $p_{k}(s)$ for different values of $k$ for $d=2$ and $\beta= 1,2,4$ in Fig.~\ref{F1}. We can also calculate $R_{2}(s)$ and $Y_{2}(s)$ from Eq.~(\ref{e63}) with $p_{n-1}(s)$ from Eq.~(\ref{e110}). We will show numerical results for $R_{2}(s)$ and $Y_{2}(s)$ for $d=2$ later.

It is convenient to define an average over level spacings:
\begin{equation}\label{e112}
\langle \langle X(s) \rangle \rangle = \int_{0}^{\infty}e^{-(2\beta+1)s}s^{\beta}X(s)ds.
\end{equation}
The variance of the nearest-neighbor spacing distribution is given by 
\begin{equation}\label{e113}
\sigma^{2}(0)= \langle \langle s^{2} f_{\beta}(s)^{2}\rangle \rangle -1.
\end{equation}
The covariance between $s_{1}$ and $s_{n}$ can be derived from Eq.~(\ref{e108}) as
\begin{equation}\label{e114}
\langle \langle s_{1}s_{n}\rangle \rangle -1 = \sum_{\mu=0}^{\beta -1}\left(\frac{\lambda_{\mu}}{\lambda_{\beta}}\right)^{n-1} \langle \langle sf_{\beta}f_{\mu}\rangle \rangle^{2},
\end{equation}
valid for $n\geq 2$. Note that the covariance goes to zero exponentially as $n$ increases. 

Using Eqs.~(\ref{e113}) and (\ref{e114}) in Eq.~(\ref{e62}), we obtain the variance of the $n^{\rm th}$ spacing distribution as
\begin{align}\label{e115}
\sigma^{2}(n-1)=& n\sigma^{2}(0)+\sum_{j=2}^{n}(n-j+1)\langle \langle s_{1}s_{j}\rangle \rangle\nonumber\\
=&n\left[\langle \langle s^{2}f_{\beta}^{2}\rangle \rangle -1 + 2\sum_{\mu=0}^{\beta -1}  \frac{\lambda_{\mu}}{\lambda_{\beta}-\lambda_{\mu}} \langle \langle sf_{\beta}f_{\mu}\rangle \rangle^{2}\right]-2\sum_{\mu=0}^{\beta-1}\frac{\lambda_{\mu}\lambda_{\beta}}{(\lambda_{\beta}-\lambda_{\mu})^{2}} \langle \langle sf_{\beta}f_{\mu}\rangle \rangle^{2}\nonumber\\
&+ 2\sum_{\mu=0}^{\beta-1}\left(\frac{\lambda_{\mu}}{\lambda_{\beta}} \right)^{n}\frac{\lambda_{\mu}\lambda_{\beta}}{(\lambda_{\beta}-\lambda_{\mu})^{2}} \langle \langle sf_{\beta}f_{\mu}\rangle \rangle^2.
\end{align}
The result in Eq.~(\ref{e115}) consists of three terms: linear in $n$, constant, and exponentially decaying in $n$.
In the large-$n$ limit, the spacing variance becomes
\begin{equation}\label{e116}
\sigma^{2}(n-1)= \frac{n}{2\beta+1} + \gamma(\beta,2),
\end{equation}
where $\gamma (\beta,2)$ is a constant. This will be explicitly verified below for $\beta=1,2,4$. We will use the notation $\gamma(\beta, d)$ to denote this constant for general $d$.

For $\beta=1$, one can derive the above quantities explicitly. The two eigenvalues and their eigenfunctions are as follows:
\begin{align}
\lambda_{0}&=\left(1-\sqrt\frac{3}{2}\right)\frac{2}{27}, & \lambda_{1}&=\left(1+\sqrt\frac{3}{2}\right)\frac{2}{27}, \label{e117} \\
f_{0}(s)&= \frac{\left(s-\sqrt{\frac{2}{3}} \right)}{\sqrt{2|\lambda_{0}|\sqrt{\frac{2}{3}}}},& f_{1}(s)&= \frac{\left(s+\sqrt{\frac{2}{3}}\right)}{\sqrt{2|\lambda_{1}|\sqrt{\frac{2}{3}}}}. \label{e118}
\end{align}
The other statistical quantities are 
\begin{align}
&\sigma^{2}(0)= \frac{1}{3}\left(2-\sqrt{\frac{2}{3}}\right),\label{e119}\\
&\langle \langle s_{1}s_{n}\rangle \rangle -1 = \frac{(-1)^{n-1}}{3}\frac{\left(\sqrt{\frac{3}{2}}-1 \right)^{n-1}}{\left(\sqrt{\frac{3}{2}}+1 \right)^{n-1}}~, \label{e120}
\end{align}
and
\begin{align}
&\sigma^{2}(n-1)= \frac{n}{3} + \frac{1}{18} + \frac{(-1)^{n-1}}{18}\frac{\left(\sqrt{\frac{3}{2}}-1 \right)^{n}}{\left(\sqrt{\frac{3}{2}}+1 \right)^{n}}\rightarrow\frac{n}{3}+\frac{1}{18} \quad \mbox{for large}~n. \label{e121}
\end{align}
The equalities in Eqs.~(\ref{e120}) and (\ref{e121}) apply for $n \geq 2$.

For $\beta=2,4$ we have numerically computed the eigenvalues and eigenfunctions from the integral equation (\ref{e101}). These are tabulated in the Appendix of this paper. Figs.~\ref{F1}-\ref{F6} show different statistical measures obtained from numerical integrations for $d=2$ and $\beta=1,2,4$. In Fig.~\ref{F1}, we show the higher-order spacing distributions $p_k(s)$ for $\beta=1,2,4$. In Fig.~\ref{F2}, we make a comparison between the spacing distributions $p_0, p_1, p_2$ for all three $\beta$-values. In Fig.~\ref{F3}, we show a comparison of $p_0(s)$ for $d=0,1,2$ for each $\beta$. In Figs.~\ref{F3}(a), (b), (c), we have also included the corresponding classical-ensemble results for GOE, GUE and GSE, respectively. (As mentioned earlier, these arise for $d=N-1$.) These plots demonstrate the crossover from Poisson to classical results in the FRCG models as $d$ is increased from 0 to $N-1$. In Fig.~\ref{F4}, we show a comparison between our exact results and the MF approximation. We have compared results for $p_{k_m}(s)$ vs. $s$, where $k_m$ is the smallest value for which the exact results are numerically indistinguishable from the MF results on this scale. The quality of agreement improves further for higher values of $k$. Notice that $k_m$ is smaller for higher values of $\beta$. Fig.~\ref{F5} and Fig.~\ref{F6} correspond to the two-point correlation function $R_2(s)$ and the two-point cluster function $Y_2(s)$, respectively.

We have also used the eigenvalues and eigenfunctions to compute $\sigma^2(n-1)$ and verified Eq.~(\ref{e116}) with $\gamma(2,2)= 0.0451566$ and $\gamma(4,2)=0.0305985$. For $\beta=1,\gamma(1,2) = 1/18$ as in Eq.~(\ref{e121}).

Following the above discussion regarding $\sigma^2(n-1)$, and from the result for $\Sigma^{2}(n)$ in the $d=1$ case in Eq.~(\ref{e87}), we rewrite the result of number variance for $d=2$ as 
\begin{equation}\label{e122}
\Sigma^{2}(n)= \sigma^{2}(n-1)+ \frac{2\beta(2\beta+2)}{6(2\beta+1)^{2}}.
\end{equation}
This will be verified by MC calculations in paper II.

\section{Fluctuation Properties for $d>2$}
\label{s11}

The $d> 2$ case is analytically less tractable, as expected. However, the basic structure of many of the equations is formally similar to that for the $d=2$ case. Let us start with the jpd in Eq.~(\ref{e55}) for $d> 2$. We follow the calculations in Sec.~\ref{s10}. The first major change is encountered in the integral equation (\ref{e101}), which now becomes
\begin{align} \label{e123}
\int_{0}^{\infty}e^{-(\beta d+1)t}t^{\beta}(t+s_{d-1})^{\beta}\cdots(t+\cdots+s_{1})^{\beta} \times \nonumber\\
f_{\mu}(t,s_{d-1},\cdots,s_{2})dt=\lambda_{\mu}f_{\mu}(s_{d-1},\cdots,s_{1}).
\end{align}
In this case, there are $\zeta+1$ eigenvalues and eigenfunctions, where $\zeta$ can be calculated from the multinomial form of $f_\mu$:
\begin{equation}
\zeta = (\beta+1)(3\beta+1) \cdots \left( \frac{d(d-1)}{2} \beta + 1 \right) .
\end{equation}
Thus $\mu= 0,1,\cdots,\zeta$. We consider the eigenvalues in ascending order with $\lambda_{\zeta}$ as the highest.

The appropriate generalization of Eq.~(\ref{e102}) is 
\begin{align}\label{e124}
&P_{d}^{(n)}(s_{1},\cdots,s_{n})=\frac{1}{\lambda_{\zeta}^{n-1}}\exp\left[-(\beta d+1)\sum_{i=1}^{n}s_{j}\right] \times \nonumber \\
&\left\lbrace \prod_{k=0}^{d-1}\prod_{j=1}^{(n-k,1)_{>}}(s_{j}+s_{j+1}+ \cdots+s_{j+k})^{\beta} \right\rbrace 
f_{\zeta}(s_{1},\cdots,s_{d-1})f_{\zeta}(s_{n},\cdots,s_{n-d+2}).
\end{align}
This expression can be used to calculate various fluctuation properties. However, the subsequent calculations become more involved as $d$ increases. Thus, we use MC techniques (described in paper II) to obtain the complete picture for arbitrary $d$.

As we have mentioned in the introductory section, significant work on FRCG models was also done by Bogomolny et al. (BGS) \cite{bgs99,bgs01}. It is useful to compare and contrast our approach with that of BGS. The cases $d=0,1$ are simple, so we will focus on cases with $d \ge 2$. \\
1) We use the integral equation to directly calculate the jpds and the corresponding fluctuation properties. In this context, we adopt a scaling approach to simplify the complicated integral equations. This should be contrasted with Ref.~\cite{bgs01}, where BGS use a saddle-point approximation to make the equations tractable. \\
2) Due to the simplicity of our approach, we are able to explicitly calculate important physical quantities (inclusive of prefactors), e.g., level densities, spacing variance, number variance, spacing distributions, etc. \\
3) We present a MF approximation for $d \geq 2$, which enables the straightforward calculation of all statistical quantities for arbitrary $d$. Our MF approximation is validated by comparison with analytical results for $d=2$, and MC results for $d > 2$ (presented in paper II). \\
4) In paper II, we present detailed MC results for the case with arbitrary $d$, enabling us to characterize the crossover from the Poisson limit to the Wigner-Dyson limit. \\
5) We present detailed numerical results in paper II to demonstrate that FRCG models provide a good framework to understand the spectral statistics of QKR and BRM. \\
6) In the context of point 5), we formulate FRCG models with fractional $d$ to cover the entire parameter range for QKR and BRM.

\section{Summary and Discussion}
\label{s12}

In this paper, we have generalized Dyson's Brownian motion model for eigenvalues of random matrix ensembles to introduce finite-range Coulomb gas (FRCG) models. These FRCG models are parametrized by the range $d$, which characterizes the extent of the particle-particle interaction in the Coulomb gas model. The FRCG models are solvable, and we have presented detailed analytical results. In this context, we calculate various fluctuation properties, e.g., spacing distributions, spacing variance, two-level correlation functions, etc. Further, we have presented an approximate mean-field (MF) solution which works very well for $d \geq 2$. This MF solution is validated by comparison with exact results for $d=2$.

From our analytical and numerical results, we observe that FRCG models provide an elegant route for transition from Poisson to classical ensembles as the interaction range increases from $d=0$ to $d=N-1$. Further, the onset of this transition is rapid. There are several interesting features of this crossover. For example, the number variance $\Sigma^2(r)$ exhibits a linear dependence on $r$ for FRCG models with $d=O(1)$, whereas this quantity grows logarithmically ($\sim \ln r$) for the classical ensembles. Moreover, $Y_2(s)$ decays exponentially (or faster) for small $d$, but decays algebraically for the classical ensembles. It is clearly relevant to quantify this crossover as a function of $d$, and we undertake this task in paper II of this two-part exposition.

In paper II, we present Monte Carlo (MC) results for FRCG models. The MC approach will provide ``exact'' results for FRCG models at intermediate values of $d$, where the framework described in paper I becomes unwieldy. Our results will confirm that FRCG models provide a new universality class of random matrix ensembles. In paper II, we will also study the applications of FRCG models. In this context, we will discuss banded random matrices (BRM) and quantum kicked rotors (QKR) in detail. We will also demonstrate that FRCG models are appropriate models for BRM and QKR.

\newpage
\section*{Appendix}

\subsection*{Eigenvalues and Eigenvectors for $d=2$, and $\beta=1,2,4$} 

\begin{tabular}{ |p{3cm}|p{5.5cm}|}
\hline
\multicolumn{2}{|c|}{$d=2, \beta=1$} \\
\hline
Eigenvalue ($\lambda_{\mu}$) & Eigenfunction ($f_{\mu}(s)$) \\
\hline
$\lambda_{0}$=-0.01665 & $ -4.95204+6.06499s$  \\
\hline
$\lambda_{1}$=0.16479  & $ 1.57394+1.92768s$  \\
\hline
\end{tabular}
\\
\\
\\
\begin{tabular}{ |p{3cm}|p{7cm}|}
\hline
\multicolumn{2}{|c|}{$d=2, \beta=2$} \\
\hline
Eigenvalue ($\lambda_{\mu}$) & Eigenfunction ($f_{\mu}(s)$) \\
\hline
$\lambda_{0}$=0.000243 & $ 23.23395+68.27115s+38.14544s^{2}$  \\
\hline
$\lambda_{1}$=-0.00444 & $ -8.88387-1.26202s+12.73137s^{2}$  \\
\hline
$\lambda_{2}$=0.03491  & {$-2.50206-4.99504s-3.23088s^{2}$} \\
\hline
\end{tabular}
\\
\\
\\
\begin{tabular}{ |p{3cm}|p{11cm}|}
\hline
\multicolumn{2}{|c|}{$d=2, \beta=4$} \\
\hline
Eigenvalue ($\lambda_{\mu}$) & Eigenfunction ($f_{\mu}(s)$) \\
\hline
$\lambda_{0}$=0 & $ 506.72315-3395.75746s+7145.49725s^{2}-5752.00876s^{3}+1524.93879s^{4}$  \\
\hline
$\lambda_{1}$=-0.000002 & $ 214.39577-650.10829s-71.31645s^{2}+1073.96383s^{3}-523.06543s^{4}$  \\
\hline
$\lambda_{2}$=0.00003  & {$80.95249-15.73572s-231.86297s^{2}-57.01816s^{3}+161.12807s^{4}$} \\
\hline
$\lambda_{3}$=-0.00028 & {$25.73671+50.34554s+6.04437s^{2}-55.74946s^{3}-42.18508s^{4}$} \\
\hline
$\lambda_{4}$=0.00192 & {$ -5.96673-21.16616s-31.703158s^{2}-24.21054s^{3}-8.16136s^{4}$} \\
\hline
\end{tabular}
\\
\\
\\
\\

\newpage

\begin{figure}[H]
\centering
\includegraphics[width=0.6\textwidth]{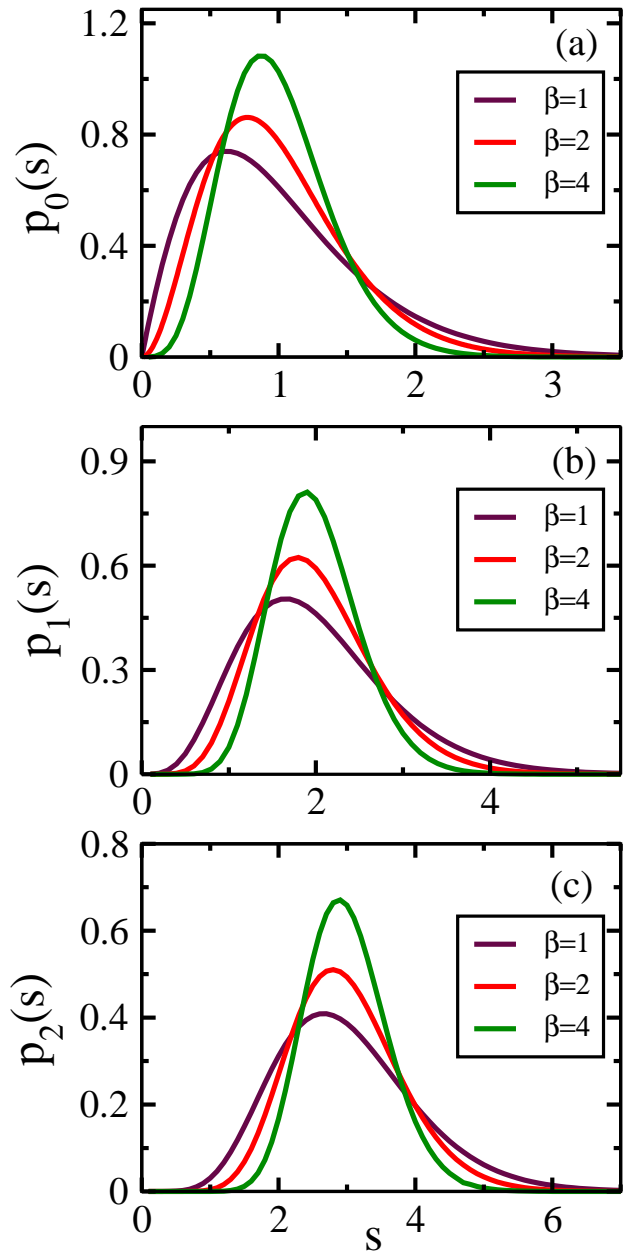}
\caption{Comparison of spacing distributions for $d=2$ and $\beta=1,2,4$. The different frames show exact results for (a) $p_0(s)$, (b) $p_1(s)$, and (c) $p_2(s)$.}\label{F2}
\end{figure}

\begin{figure}[H]
\centering
\includegraphics[width=0.6\textwidth]{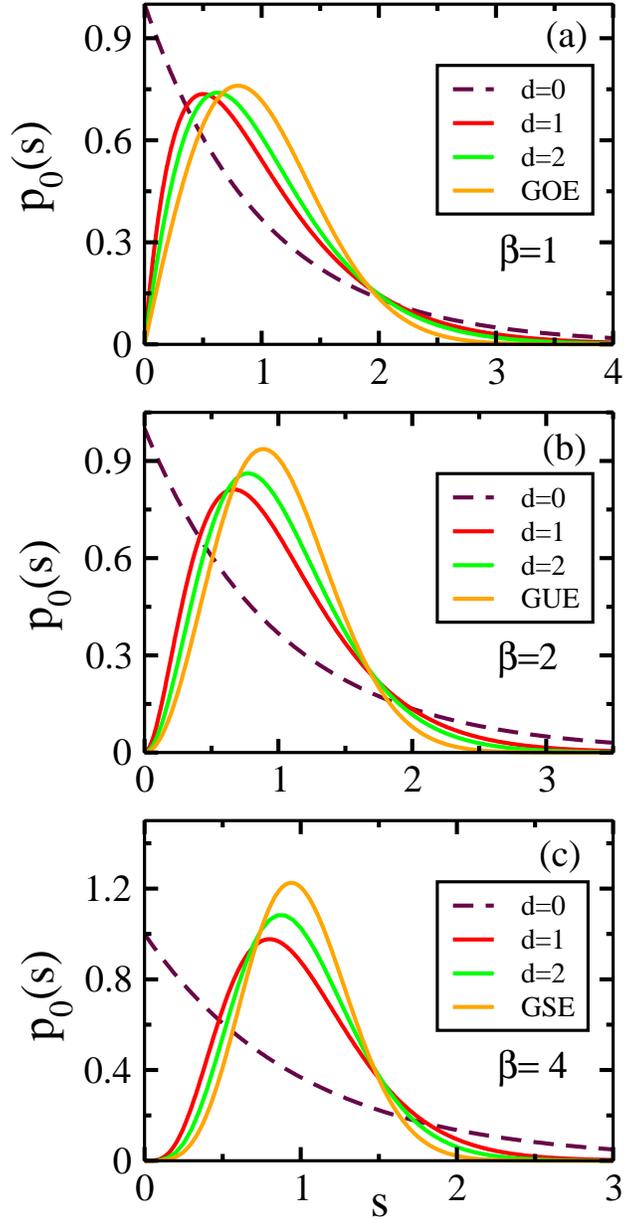}
\caption{Nearest-neighbor spacing distribution, $p_0(s)$ vs. $s$, for $d=0,1,2$ and (a) $\beta=1$, (b) $\beta=2$, and (c) $\beta=4$. The $d=0$ case corresponds to Poisson ensembles. For comparison, we also show the corresponding results for classical ensembles, which arise for $d=N-1$. These correspond to GOE, GUE and GSE for $\beta = 1,2,4$ respectively.}\label{F3}
\end{figure}

\begin{figure}[H]
\centering
\includegraphics[width=0.6\textwidth]{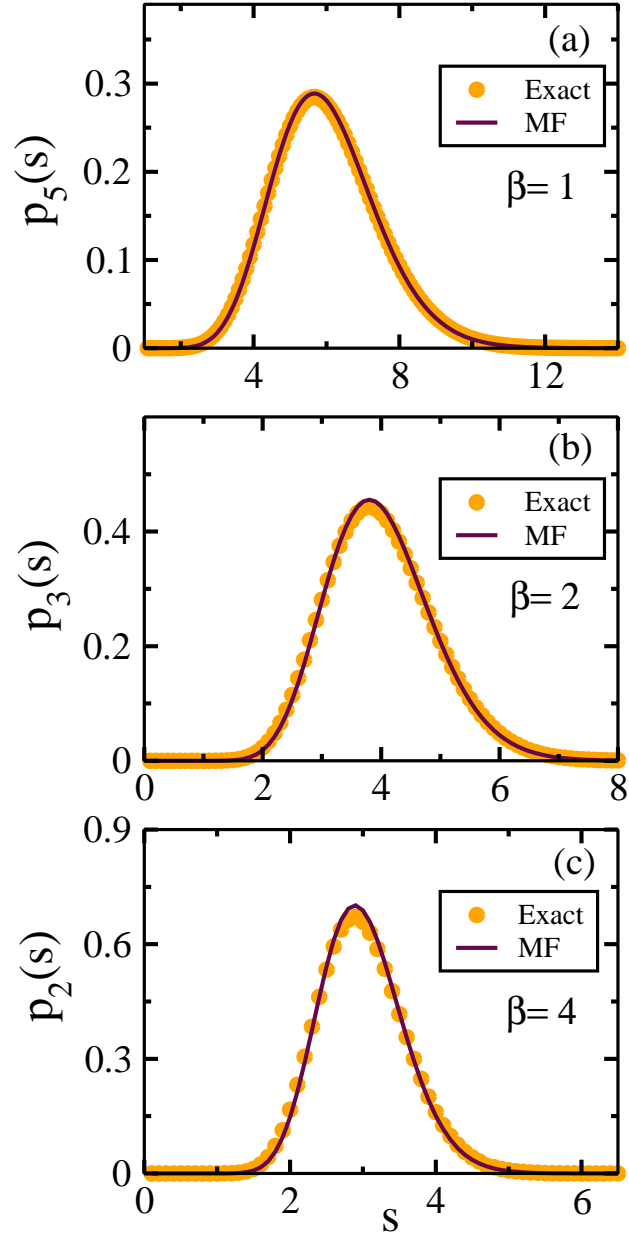}
\caption{Comparison between exact and mean-field (MF) results for $p_{k_m}(s)$ in the $d=2$ case. The different frames show the cases (a) $\beta=1$, (b) $\beta=2$, and (c) $\beta=4$. For $k \ge k_m$, the exact and MF results are numerically indistinguishable on the scale of this plot.}\label{F4}
\end{figure}

\begin{figure}[H]
\centering
\includegraphics[width=0.6\textwidth]{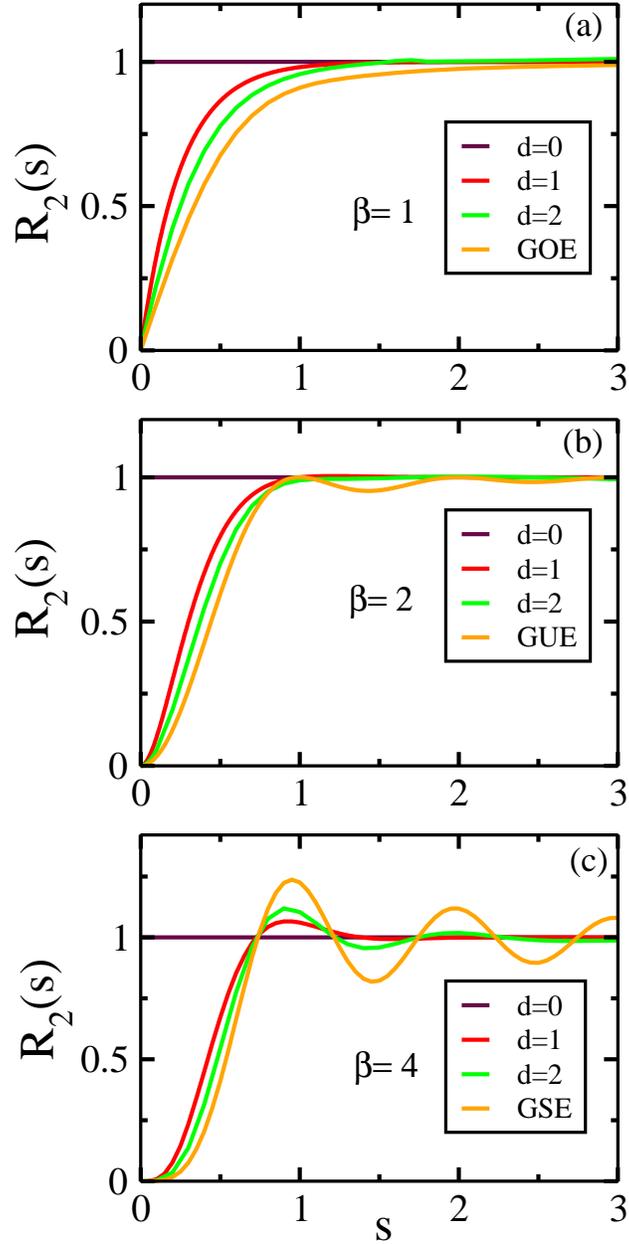}
\caption{Two-point correlation function $R_2(s)$ for (a) $\beta=1$, (b) $\beta=2$, and (c) $\beta=4$. In each frame, we plot $R_2(s)$ vs. $s$ for $d=0,1,2$ and the classical-ensemble result ($d=N-1$). The $d=0$ case yields the Poisson result.}\label{F5}
\end{figure}

\begin{figure}[H]
\centering
\includegraphics[width=0.6\textwidth]{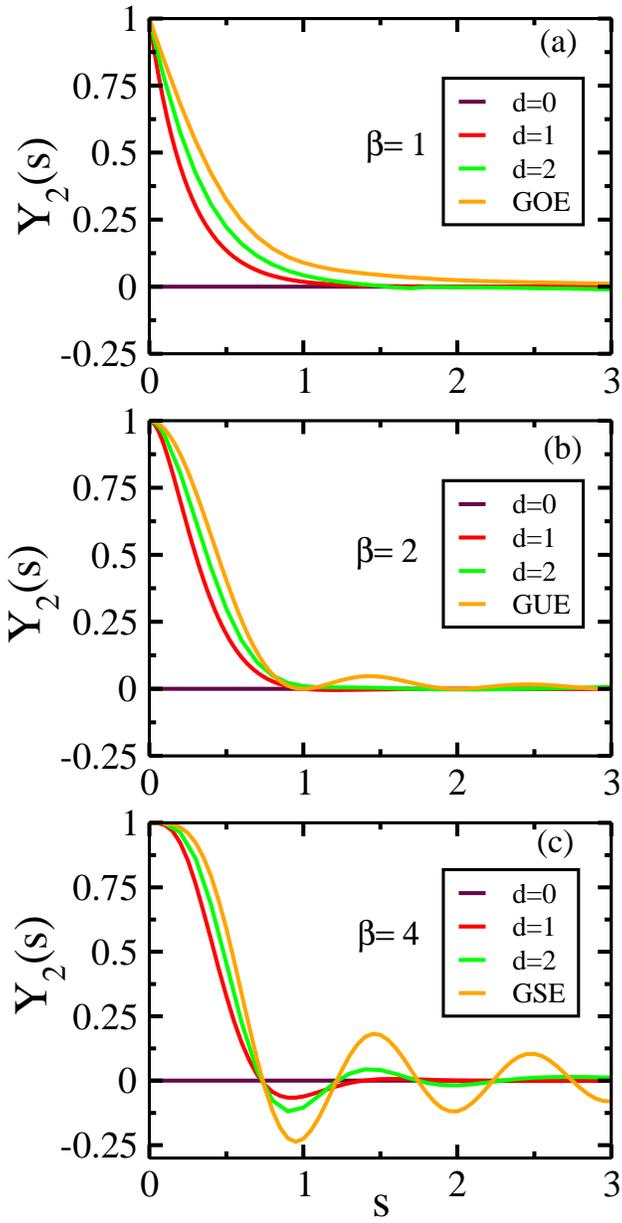}
\caption{Analogous to Fig.~\ref{F5}, but for the two-point cluster function $Y_2(s) = 1-R_2(s)$.}\label{F6}
\end{figure}

\end{document}